\documentclass[oribibl]{llncs}

\usepackage{amsmath}
\usepackage[colorlinks = true,
  linkcolor = blue,
  citecolor = blue,
  urlcolor = blue]{hyperref}

\usepackage{url}
\usepackage{natbib}
\usepackage{booktabs}
\usepackage{tabularx}
\usepackage{graphicx}
\usepackage{enumitem}

\begin{document}

\title{Early Perspectives on the \\ Digital Europe Programme}

\author{Jukka Ruohonen\inst{1}\orcidID{0000-0001-5147-3084} \and Paul Timmers\inst{2} \institute{University of Southern Denmark, \email{juk@mmmi.sdu.dk} \and KU Leuven \& University of Oxford, \email{paul.timmers@kuleuven.be}}}

\maketitle

\begin{abstract}
A new Digital Europe Programme (DEP), a funding instrument for development and
innovation, was established in the European Union (EU) in 2021. The paper makes
an empirical inquiry into the projects funded through the DEP. According to the
results, the projects align well with the DEP's strategic focus on cyber
security, artificial intelligence, high-performance computing, innovation hubs,
small- and medium-sized enterprises, and education. Most of the projects have
received an equal amount of national and EU funding. Although national origins
of participating organizations do not explain the amounts of funding granted,
there is a rather strong tendency for national organizations to primarily
collaborate with other national organizations. Finally, information about the
technological domains addressed and the economic sectors involved provides
decent explanatory power for statistically explaining the funding amounts
granted. With these results and the accompanying discussion, the paper
contributes to the timely debate about innovation, technology development, and
industrial policy in Europe.
\end{abstract}

\keywords{public policy, funding, consortia, key technologies, innovation hubs}

\section*{Introduction}

The Digital Europe Programme was established by the von der Leyen's first
Commission with the specific Regulation (EU) 2021/694. It is a large-scale
funding instrument with a focus on all and everything related to
digitalization. Unlike the large-scale Horizon Europe programme, the DEP is on
the side of development and innovation, not on the side of scientific research,
whether basic or applied. Having said that, the regulation emphasizes synergies
with other funding instruments, including not only the Horizon Europe programme
but also the Connecting Europe Facility (CEF) and the Recovery and Resilience
Facility established during the coronavirus pandemic. Despite the DEP's
generally broad scope, it has some strategic focus areas. Among them are
artificial intelligence (AI) and small- and medium-sized enterprises
(SMEs). Both are important for the Europe's current and future competitiveness
in the face of ever-increasing global competition. Many studies could be
referenced to strengthen these points, but it suffices to remark results
according to which the adoption rate of AI in both public and private sectors
has still been slow in many European countries~\citep{EC22b, Krogstie24}. At the
same time, the general degree of digitalization varies greatly across the member
states~\citep{Brodny22}. As always with technology, however, the problem is not
only about adoption and geographic cohesion. Among other things, also cyber
security remains a pressing issue throughout the EU. Nor are people's existing
skills and knowledge enough to answers to these and other grand challenges.

According to the empirical dataset soon elaborated, about one point five billion
euros have thus far been allocated from the DEP funding instrument. Given that
the DEP's overall budget is over eight billion euros, this point justifies the
wording about \textit{early} perspectives in the paper's title. This point
should be kept in mind throughout the paper; what is observed is the DEP's
initial roll-out. This point affects also the paper's overall framing and
rationale of a policy evaluation. Given a distinction between \textit{ex~ante}
appraisal and \textit{ex~post} evaluation of policies~\citep{Smismans15}, the
paper could be categorized as an early \textit{ex~post} evaluation
\textit{check}. By the noun check, it is meant that while a comprehensive
\textit{ex~post} evaluation is impossible due the DEP's ongoing nature, an early
assertion can be done akin to the concept of quality gates sometimes used in
software engineering~\citep[cf.][]{Flohr08}. Even an early check provides
valuable information regarding whether the DEP is on the right track with
respect to the \textit{ex~ante} priorities and strategies specified for it. To
this end, the paper's general rationale is similar to the midterm or interim
evaluations commonly conducted for the EU's research and development
programs. As for evaluation research in general and its
taxonomies~\citep{Schoenefeld17}, the paper follows a \textit{hierarchical} and
\textit{informal} evaluation approach. By the latter adjective, it is meant that
no strict protocol or standard is used for the empirical assessment. Regarding
the former adjective as well as the criteria for the early or preliminary
evaluation, the paper takes a bottom-up approach by focusing on the projects
funded through the DEP and then assessing whether and how well they have matched
the high-level objectives specified in Regulation (EU) 2021/694.

Without further delay, the following research questions (RQs) are examined:
\begin{enumerate}[label=RQ$_\arabic{enumi}$]
\itemsep 5pt
\item{What technological domains have been addressed in the DEP-funded projects
  and which economic sectors have been involved in the projects?}
\item{How much national funding has been granted alongside the EU funding?}
\item{How large have the consortia been behind the DEP-funded projects and what
  have their geographic composition been?}
\item{Do the technological domains explain the amounts of funding granted?}
\end{enumerate}

\enlargethispage{2cm} 

There are two sections upfront before the empirical examination of these four
research questions. The first section is a brief but sufficient take for better
motivating the RQs and the paper's topic in general. It also better connects the
paper to existing research as well as the timely political debates in the
EU. The subsequent section presents the dataset and the methods for examining
it. Presentation of the empirical results follows. The final concluding section
summarizes the answers to the four research questions alongside presenting a
theoretical reflection and some points regarding further research paths paved.

\section*{Background and Related Research}

The EU's industrial policy and funding for it have seen a large transformation
in recent years. Due to recurring crises, starting from the 2008 global
financial crisis and going through the migration crisis in 2015, the coronavirus
pandemic in 2020, and more recently the war in Ukraine, not forgetting the
existential climate change crisis, the EU has been forced to adapt its financing
policies. In terms of industrial policy, there has been a visible move from the
EU's traditional cohesion policy toward funding sectors and technologies that
are perceived as critical~\citep{Molica25}. At the same time, strings have been
loosened for direct state aid to national industries and so-called national
champions by the member states. In many ways, these developments strike in the
heart of the EU because they affect the union's celebrated internal market and
its functioning.

In recent years, many visions have been presented and proposed to improve the
functioning of the internal market. Among other things, \citet{Letta24} proposed
that the EU's treaties should be altered by including a new freedom to research,
innovate, and develop. Then, the \citeauthor{Draghi24}'s \citeyearpar{Draghi24}
recent strategy vision for the EU's competitiveness recommends further large
joint public investments together with improvements to the union's innovation
capacity. By and large, the report builds upon ideas that large-scale public
funding, not private sector investments, should drive the union's innovation and
industrial policies, including scaling of European companies and the financing
of technologies and their commercialization~\citep{CESinfo24}. ``To unleash
private investment, we also need public funding'', stated \citet{vonderLeyen24}
in her recent talk at a plenary session of the European Parliament. Although
much smaller than what is discussed and perhaps being already planned, also the
DEP can be framed against this line of thought about public funding. This
political background justifies also the paper's relevance and timeliness. It is
important to know more about the DEP even when keeping in mind its ongoing
nature.

The concept of differentiation commonly used in the policy area of European
security and defense provides a high-level theoretical lens to guide the
empirical analysis. Although there are no universally agreed definitions for the
concept, it is essentially about working together in non-homogenous and flexible
ways that include but go beyond instances of integration and pooling of
sovereignty at the EU-level~\citep{Siddi22}. In terms of European integration,
differentiation has involved different opt-in and opt-out choices provided to
the member states with respect to particular policy questions and domains they
have, or have not, perceived as relevant for accelerating
integration~\citep{Rieker24}. In other words, instead of the older idea about
wholesale integration of the whole Europe, differentiated integration has
provided ways for groups of member states to deepen integration in some
particularly policy areas.

When translated to the domain of innovation, technology development, and
industrial policy, differentiation can be seen already in the establishment of
new funding instruments for specific policy needs. Even when considering the
CEF, DEP, and Horizon Europe alone, the EU has established over ten new joint
undertakings, including those related to high-performance computing, cyber
security, railways, the green transition, and healthcare~\citep{ECA24}. While
these have been driven through the EU-level, they have often provided different
opt-in choices for the member states to engage further. Also the DEP builds upon
a similar differentiation rationale in a sense that also national funding is
typically required for obtaining EU funding through the instrument. When
compared to the traditional rationale for cohesion policy, differentiation can
be seen to work also underneath the selective or smart specialization approach
seeking to prioritize support for particular sectors, industries, companies, and
key technologies~\citep{Doussineau21, Molica25}. Differentiation is a useful
theoretical concept also because it can be used to separate the DEP from science
funding available through the much larger Horizon Europe programme.

In terms of further related work, the paper aligns with the existing and
extensive research on European innovation ecosystems and their funding
instruments, including the empirical research on the Horizon Europe
programme~\citep[among many others]{Fernandez19, Kosztyan24,
  Veugelers15}. However, to the best of the author's knowledge, there are no
explicitly related works on the DEP specifically. Sure enough, the DEP is
frequently mentioned in the recent innovation
literature~\citep[e.g.,][]{Colovic25}, but, thus far, there seems to be no
existing works that would have taken an in-depth empirical look at the programme
and the projects it has funded. Thus, in addition to the timeliness and
political relevance, the paper addresses a gap in existing knowledge. By helping
at covering the knowledge gap, a few relevant insights are available also
regarding the theorization with the differentiation concept.

\section*{Data and Methods}

The data was collected in December 2024 from the EU's online portal for funding
and tenders~\citep{EC24a}. During that time, there were $492$ projects funded
through the DEP, although one of these had to be excluded because no monetary
amounts were available for it. The clear majority of projects were still ongoing
during the data collection---in fact, only about 8\% had ended.

For each project, the total budget and the amount of EU funding received were
recorded. In addition, a project's description was used to categorize the
technological domains addressed and the economic sectors involved in the
project. This categorization was based on qualitative content analysis, which is
a classical and likely the simplest qualitative method available. In essence, it
seeks to iteratively and inductively find key concepts, patterns, or categories
from textual data. The choice of this simple qualitative method can be justified
on the grounds that the project descriptions are very short; typically, only a
single paragraph is available in the online portal for each project. To this
end, the focus is also on manifest content, not latent content obtainable
through abstraction and interpretation, as is often common in thematic
analysis~\citep{Lindgren20, Vaismoradi19}. In terms of coding, existing research
was followed by initially identifying many domains and sectors, and then
collating and merging overlapping and outlying domains and
sectors~\citep{Ruohonen24JSS}. After the iterations and collations, eleven
technological domains and twenty economic sectors were identified. These are
enumerated in Table~\ref{tab: domains and sectors}.

\begin{table}[th!b]
\centering
\caption{Technological Domains and Economic Sectors}
\label{tab: domains and sectors}
\begin{tabularx}{\linewidth}{lXclX}
\toprule
& Technological domain &\qquad\qquad&& Economic Sector \\
\hline
1. & AI, high-performance computing (HPC), and cloud computing && 1. & Agriculture (incl.~forestry) \\
\cmidrule{2-2}\cmidrule{5-5}
2. & Augmented reality (AR), virtual reality (VR), and three-dimensional (3D) applications && 2. & Aviation \\
\cmidrule{2-2}\cmidrule{5-5}
3. & Blockchains && 3. & Commerce, retail, and services \\
\cmidrule{2-2}\cmidrule{5-5}
4. & Cyber security && 4. & Construction \\
\cmidrule{2-2}\cmidrule{5-5}
5. & Data spaces && 5. & Culture \\
\cmidrule{2-2}\cmidrule{5-5}
6. & Digital twins && 6. & Defense \\
\cmidrule{2-2}\cmidrule{5-5}
7. & Internet of things (IoT) && 7. & Education \\
\cmidrule{2-2}\cmidrule{5-5}
8. & Quantum computing && 8. & Energy and the green transition \\
\cmidrule{2-2}\cmidrule{5-5}
9. & Robotics (incl.~all other cyber-physical systems, such as drones) && 9. & Finance and banking \\
\cmidrule{2-2}\cmidrule{5-5}
10. & Smart cities (incl.~smart housing) && 10. & Healthcare \\
\cmidrule{2-2}\cmidrule{5-5}
11. & Telecommunications (excl.~IoT) && 11. & Hubs \\
\cmidrule{5-5}
&&& 12. & Manufacturing \\
\cmidrule{5-5}
&&& 13. & Maritime (incl.~drinking and waste water systems) \\
\cmidrule{5-5}
&&& 14. & Media \\
\cmidrule{5-5}
&&& 15. & Public administration \\
\cmidrule{5-5}
&&& 16. & Safety for children \\
\cmidrule{5-5}
&&& 17. & Semiconductors (incl.~microprocessors and all other chips) \\
\cmidrule{5-5}
&&& 18. & Small- and medium-sized enterprises \\
\cmidrule{5-5}
&&& 19. & Tourism \\
\cmidrule{5-5}
&&& 20. & Transport, mobility, and logistics \\
\bottomrule
\end{tabularx}
\end{table}

The technological domains were relatively straightforward to identify, but some
concessions had to be made with the economic sectors. For instance, the
manufacturing sector was collated to contain all economic activities involved in
production of goods; hence, it contains also chemical and textile industries,
goods produced through craftsmanship, and anything labeled as ``industry 4.0''
or ``industry 5.0''. Analogously, the healthcare sector was collated to contain
all biomedical applications and pharmaceuticals. A further point is that in a
few cases the technological domains and sectors intersect. For instance, the
defense sector contains dual-use technologies, while the public administration
sector contains all ``government technologies'' (or ``govtech'', as sometimes
abbreviated in the portal), whether information systems for social welfare or
law enforcement.

In addition to the $n = 491$ projects observed, data was collected on all
organizations, whether universities, associations, or companies, who had taken
part in consortia behind the projects. This organizational level data contains
$m = 5,556$ observations, meaning that on average about eleven organizations
took part in DEP-funded consortia. To simplify the forthcoming analysis,
however, the organizations are observed through their national origins in the EU
or, in a few cases, elsewhere. These national origins are provided by the
portal. In addition, it provides data on the amount of EU funds granted for a
given organization.

Descriptive statistics and regression analysis are used to convey the
results. In terms of the former, two graphs (networks) were constructed to help
at visualization and associated analytical reasoning. The vertices in the first
project-level graph refer to the domains and sectors in Table~\ref{tab: domains
  and sectors}. Then: if a project established an innovation hub that addressed
cyber security in the manufacturing sector, there is an edge between hubs and
cyber security, an edge between hubs and manufacturing, and an edge between
cyber security and manufacturing. The edges are also weighted; if ten projects
addressed cyber security in the manufacturing sector, there is an edge with a
weight of ten between cyber security and manufacturing. The second
organization-level graph was constructed analogously by connecting the national
origins of the participating organizations. To briefly examine graph clustering
(or community detection, as it is also called in the graph context), the
algorithms of \citet{Newman06}, \citet{NewmanGirvan04}, and \citet{Rosvall07}
are used, as implemented in the \textit{igraph} package used for
computation~\citep{igraph}. Furthermore, three ordinary least squares
regressions are estimated to determine whether the monetary amounts can be
predicted by the information available. The first two regressions operate at the
project-level; the dependent variables are the logarithms of the projects' total
budgets and EU funds granted, and the primary independent variables are $31$
dummy variables representing the domains and economic sectors identified through
the qualitative content analysis. In addition, a single variable with a
continuous scale is present; it refers to the number of organizations having
taken part in consortia behind the projects. The third regression operates at
the organizational level. As some organizations have not received any EU
funding, the dependent variable is a logarithm of EU funds granted in addition
to one euro. In addition to the domains and sectors, this regression contains
additional $45$ dummy variables representing the national origins of the
organizations.

Finally, a brief note about presentation: to deepen the depth of the qualitative
content analysis, a few illuminating projects are pinpointed. To maintain rigor
and transparency, these projects are identified by their project identification
numbers, as available through the portal. Footnotes are used for bookkeeping.

\section*{Results}

\subsection*{Domains and Sectors}

The technological domains and economic sectors are a good way to start the
dissemination of the empirical results. Thus, Fig.~\ref{fig: domains and
  sectors} displays the relative percentage shares of the domains and sectors in
terms of all projects observed. As can be seen, over thirty percent of the
projects have addressed cyber security, innovation hubs, SMEs, AI (together with
HPC and cloud computing), and education. This observation is well in line with
Regulation (EU) 2021/694; among the objectives specified in Article~3(2) are
HPC, AI, cyber security and trust, and advanced digital skills. The clarifying
Articles~4, 5, 6, and 7 all also emphasize SMEs.

\begin{figure}[th!b]
\centering \includegraphics[width=\linewidth, height=8cm]{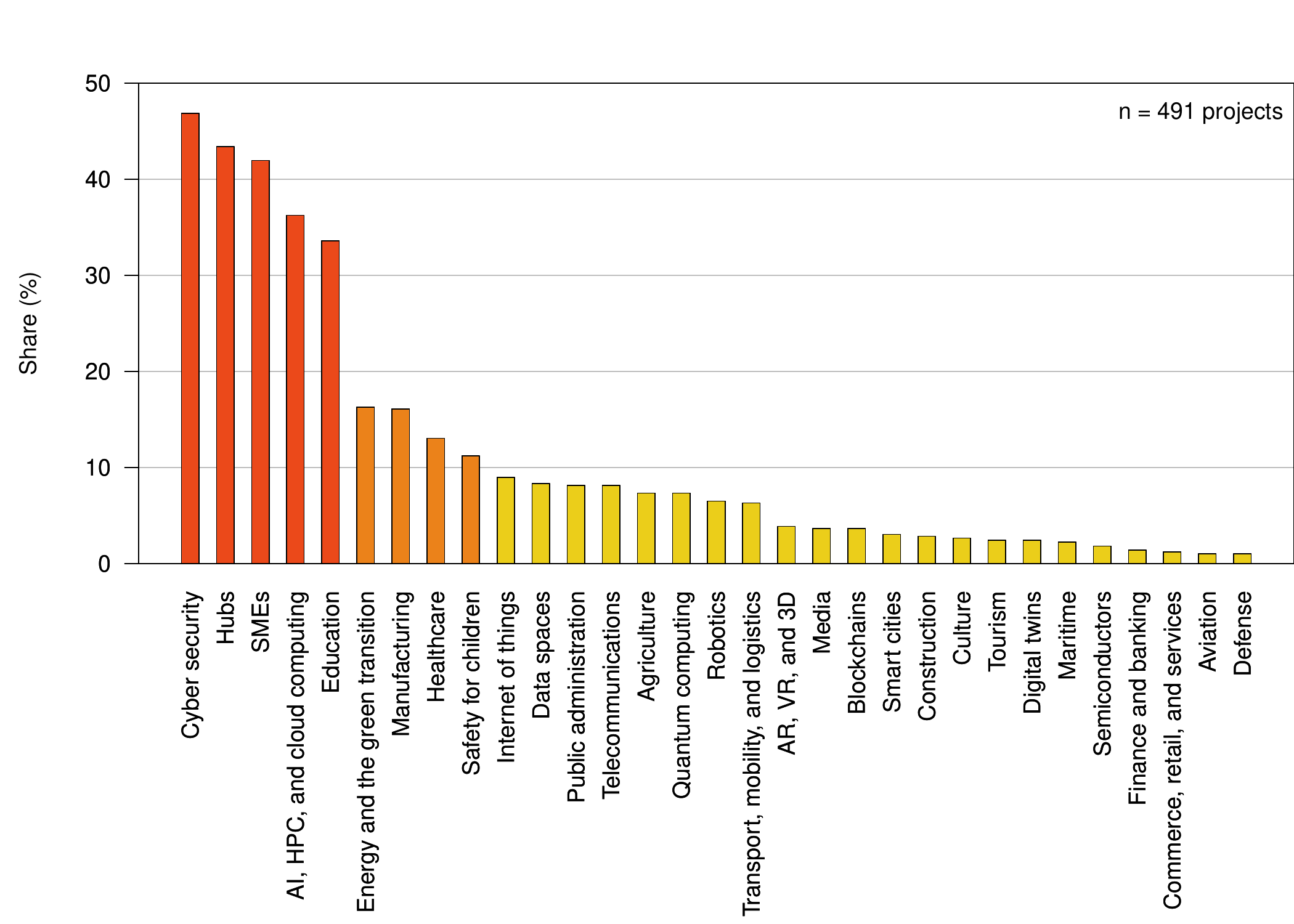}
\caption{Technological Domains and Economic Sectors}
\label{fig: domains and sectors}
\end{figure}

As also emphasized in the regulation, the objectives are in many cases
overlapping and intervening. This point applies also to the leading
technological domain, cyber security. In terms of technology \textit{per~se},
many of the cyber security projects are about developing and innovating
so-called security operations centers (SOCs), which are essentially threat
intelligence platforms. Some of the SOCs are smaller, being developed for a
deployment in a single organization, whether a company or a national,
authoritative computer security incident response team (CSIRT). Whether small or
large, many SOCs are also developed in conjunction with automation, anomaly
detection, and other AI-related techniques. Three additional points can be
raised about SOCs. Although SOCs may be developed also for commercial purposes,
the first point is that in the public sector context many ``SOCs and CSIRTs
build their capabilities in this area independently, leading to a fragmented
approach and duplication of work''.\footnote{~101128030.} Against this
illuminating quotation, a point can be raised that perhaps projects for
interoperability, standardization, and related aspects should have been funded
beforehand. Having said that, interoperability is implicitly addressed in some
of the projects developing large cross-border SOCs. The second point follows: at
least according to the project descriptions, it seems that none of the larger
SOC projects have fully considered the difficult trust-related aspects behind
cross-border sharing of threat intelligence and relate cyber security
data~\citep{Ruohonen24I3E, Serini24}. The third point is that SOCs are an
important---if not the most important---part in the law proposal for a so-called
Cyber Solidarity Act (CSA), identified as COM(2023) 209 final. For a reason or
another, however, SOCs were renamed as cyber hubs in the final compromise text
for the CSA regulation proposed~\citep{Dewi24}. The cyber security domain is
relevant also in terms of other regulations enacted in the EU.

Regarding the EU's recent cyber security regulations enacted, particularly
relevant in the present context is Directive (EU) 2022/2555, which is commonly
known as the NIS2 directive. It is a comprehensive law addressing particularly
the cyber security of Europe's critical infrastructures. It also enlarges the
scope of such infrastructures substantially from an older directive that only
considered two sectors as critical. Already because of this expansion, it is
understandable that there are well-over ten projects fully or partially dealing
with the NIS2 directive. Some of these projects are about national
transpositions of the directive, including not only in terms of jurisprudence
but also in terms of education and awareness raising among
stakeholders.\footnote{~101127928.} Others are about technical, SOC-related
implementations.\footnote{~101128069.} There are also many sectoral
projects. With or without explicit references to the NIS2 directive, examples
include the energy sector, the maritime and water sector, the agriculture sector
and food production in general, and the healthcare sector.\footnote{~101158535,
101158808, 101127567, 101127970, 101128047, and 101101522.} In this regard, a
point can be raised that perhaps the directive's new requirements could be
better---or at least more \text{inexpensively---addressed} through a
comprehensive multi-sectoral approach, as also done in one
project.\footnote{~101145879.} All this said, the NIS2 directive is not the only
one appearing in the project descriptions. Among other laws, the recently agreed
Regulation (EU) 2024/2847, the so-called Cyber Resilience Act (CRA), also
appears in the descriptions. For instance, there is a project supporting the new
standardization work imposed by the CRA.\footnote{~101158521.} Another project is
developing a compliance wizard for helping SMEs to comply with the CRA in the
future.\footnote{~101158539.}

\begin{figure}[th!b]
\centering
\includegraphics[width=\linewidth, height=10cm]{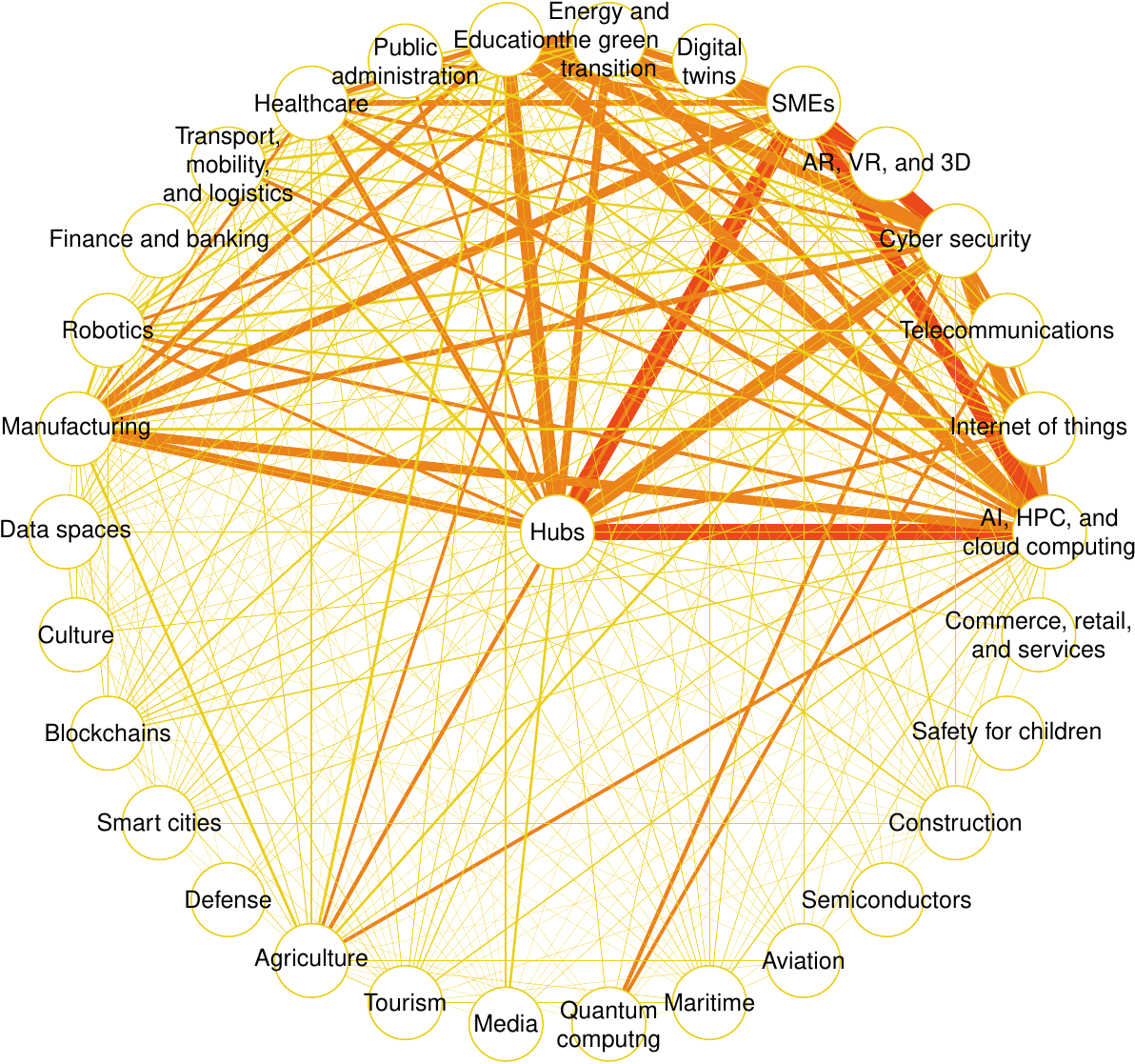}
\caption{A Graph Representation of the Technological Domains and Economic Sectors}
\label{fig: topics graph}
\end{figure}

A further particularly important cyber security law in the present context is
Regulation (EU) 2021/887, which established the new EU-level European
Cybersecurity Competence Centre (ECCC) that coordinates activities of national
coordination centers (NCCs) the member states were obliged to
establish.\footnote{~\url{https://cybersecurity-centre.europa.eu/}} According to
the regulation's Article 7, the national NCCs are tasked to improve synergies in
the EU, help at cyber security education, and facilitate development and
innovation in the domain, among other things. Against this background, it is
again no wonder that there are well-over ten projects explicitly or implicitly
dealing with the establishment of national NCCs and associated work related to
the centers. The point is important because at the same time the DEP-specific
Regulation (EU) 2021/694 specified in its Article~16(1) that European digital
innovation hubs should be established in each member state during the first year
of the funding instrument's implementation. This objective has been
satisfied. Innovation hubs, including endeavors with alternative names, whether
ecosystems or collaboration networks, take the second place in Fig.~\ref{fig:
  domains and sectors}. Some member states have also established more than one
hub. Thus, a critical point can be raised about duplication between the
innovation hubs and the NCCs, both of which more or less operate under the same
rationale and seek the same goals. Given that there were well-over five hundred
innovation hubs in Europe already before the DEP's enactment~\citep{Teixeira22},
a further related critical point is that perhaps something else might be
considered instead of merely establishing more and more hubs. Synergies would be
a good example.

Before continuing further, it should be emphasized that the cyber security
projects funded by the DEP are not only about SOCs and new EU regulations. A few
concrete examples serve to make the point explicit; there are also projects
related to penetration testing, chief security officers provided as a service,
nationwide bug bounties, cyber insurance, and secure software
development.\footnote{~101127307, 101127962, 101128013, 101128070, and
101158471.}

The project descriptions for the hub-establishing projects align well with the
\citeauthor{EC21}'s \citeyearpar{EC21} strategy documents according to which the
national innovation hubs should be one-stop-shops with a not-for-profit motive
of improving the digitalization particularly among SMEs and public sector
organizations. Hubs also have a central position in the project-level graph
representation visualized in Fig.~\ref{fig: topics graph}. Given that the widths
and colors of the edges in the visualized align with the edge weights, it can be
concluded that hubs are particularly strongly connected to SMEs, the AI/HPC
domain (including cloud computing), cyber security, education, and
manufacturing, respectively. The graph is also relatively dense. Its
density---the ratio of observed edges to all potential edges---is $0.68$. In
other words, the majority of the technological domains and economic sectors are
connected to each other, although the edge weights indicate substantial variance
in terms of the strengths of the connections. The graph's density is also seen
in the clustering results. When computed with edge weights, the algorithms of
\citet{Newman06}, \citet{NewmanGirvan04}, and \citet{Rosvall07} result three
clusters, ten clusters, and one cluster, respectively. Although there are no
right or wrong answers, the middle one with ten clusters seems suitable for
pinpointing the less well connected domains and sectors. Accordingly, there are
six clusters with just one vertex: commerce (including retail and services)
defense, media, quantum computing, semiconductors, and online safety for
children. Of these, the media sector and its own cluster deserve a brief remark:
the DEP has been used to fund several media observatories dealing with
disinformation, misinformation, and the today's general information
disorder. These observatories are also about education because they seek to
inform and educate the public about media literacy and related skills. They also
act as hubs; in addition to national collaborations, they are all associated
with the union-wide European Digital Media Observatory (EDMO)
hub.\footnote{~\url{https://edmo.eu/}} In any case, to continue with the
clustering, there are two clusters with three vertices each: the healthcare,
maritime, and tourism sectors belong to one, and the aviation, culture, and
financial and banking sectors to another. The domains and sectors belonging to a
further cluster are a little better connected; these are data spaces, digital
twins, public administration, and smart cities. All remaining domains and
sectors belong to the well-connected big cluster containing fifteen vertices.

Although contradicting the previous clustering exposition, the culture sector is
worth briefly mentioning. The reason is that it has been among the very few
sectors seeking to use AR, VR, and 3D technologies to preserve and magnify the
Europe's rich cultural heritage. This use case correlates with the establishment
of specific data spaces for cultural heritage and culture in
general.\footnote{~101100683.} Projects designing and implementing data spaces
have operated also in many other economic sectors. For instance, one project is
involved in the development of a data space in the media sector, hoping to
provide an edge against global competitors, whether in terms of news production
or entertainment.\footnote{~101123423.} A couple of illuminating data space
projects operating in the manufacturing sector are also worth pointing out. The
first project seeks to enhance dynamic asset management and predictive
maintenance in the refinery and renewable energy
industries.\footnote{~101123179.} The second one elucidates the earlier point
about trust; it seeks to build sectoral alliances in order to break data silos
that are generally prevalent in the private sector.\footnote{~101083939.} While
silos are common also in the public sector~\citep{Ruohonen24I3E}, the project's
goal is important to underline because it demonstrates that development and
innovation are not always only about new technologies. Also
the \citeauthor{EC22a}'s \citeyearpar{EC22a} strategy documents have emphasized
risks related to silos and a need to have incentive schemes to motivate
data~sharing.

The AI, HPC, and cloud computing domain is closely related to the data spaces
being developed. After all, using AI to analyze particularly large datasets
requires clouds and high-performance computing. A good example would be the
nowadays already mainstream large language models (LLMs). Regarding these
models, there is a project seeking to establish a new European hub for LLMs,
including those for languages with only a relatively few
speakers.\footnote{~101195344.} Another related project is involved in
developing a new ``AI-on-demand platform'' for SMEs and public sector
organizations.~\footnote{~101146490} Indeed, the AI/HPC domain is generally
strongly connected to SMEs also in Fig.~\ref{fig: topics graph}. However, these
examples notwithstanding, a~critical point can be raised that many---if not the
most---projects in this domain have been eager to talk about AI but without
providing any rationale or any elaboration on how they are actually planning to
use AI, why, and for what purpose. In any case, the HPC context requires some
comments too. It is closely related to the previous points. For instance, there
is a project trying to support SMEs in the use of generative AI by the means of
high-performance computing.\footnote{~101163317.} The HPC context is also
closely related to computing infrastructures, hubs, and governance structures
more generally; in Europe the EuroHPC ecosystem is particularly noteworthy in
this regard.\footnote{~\url{https://eurohpc-ju.europa.eu/}} With respect to
EuroHPC, there is a project trying to establish new competency hubs for
HPC.\footnote{~101101903.} Another related project is trying to improve support
services within the EuroHPC ecosystem.\footnote{~101139786.} Both AI and HPC are
further strongly connected to the education sector.

\subsection*{Education and Skills}

Education and skills have a strong strategic focus in the DEP---and for a good
reason. According to the \citeauthor{EC24b}'s \citeyearpar{EC24b} recent report,
an educational gap and a skill gap threaten the whole Europe's prosperity in the
long-run. According to the report, there are many union-wide problems in basic
education and basic skills, including literacy and literature skills, as well as
in the current workforce's ability and skills to tackle the challenges and
opportunities brought by digitalization, new technologies, the green transition,
and the demographic change. Nor is the higher tertiary education free of problems.

Given these challenges and the strategic focus, it is understandable that also
universities and other higher education institutions have participated in the
DEP actively, having often partnered with businesses and industry associations
to better align their curricula toward labor market demands. Many new master
programs are being developed or having been already developed thanks to funding
from the DEP. The new master-level programs include those specializing students
into AI, including its emotional variants, cyber security, data management,
digital agriculture and sustainability, digital entrepreneurship and
digitalization in business contexts, IoT and computer networking in general,
robotics, automation and digitalization in healthcare, societal aspects of
digitalization, hardware and microprocessor design, quantum computing, and
probably more.\footnote{~101083531, 101083880, 101083896, 101084013, 101100757,
101123009, 101123118, 101123258, 101123289, 101123121, 101123430, 101158828, and
101189994.} In addition, there are numerous projects seeking to improve the
skills and knowledge of current employees employed by SMEs in particular. A full
scope of digital education is being used in these projects, including everything
from massive open online courses to gamification. It remains to be seen whether
these help at tackling the challenges related to re-education and upskilling of
the contemporary workforce. Regarding basic education, basic skills, and
children, it could be also argued that digitalization is a part of the
problem in itself.

\subsection*{Budgets}

\begin{figure}[th!b]
\centering
\includegraphics[width=\linewidth, height=4.5cm]{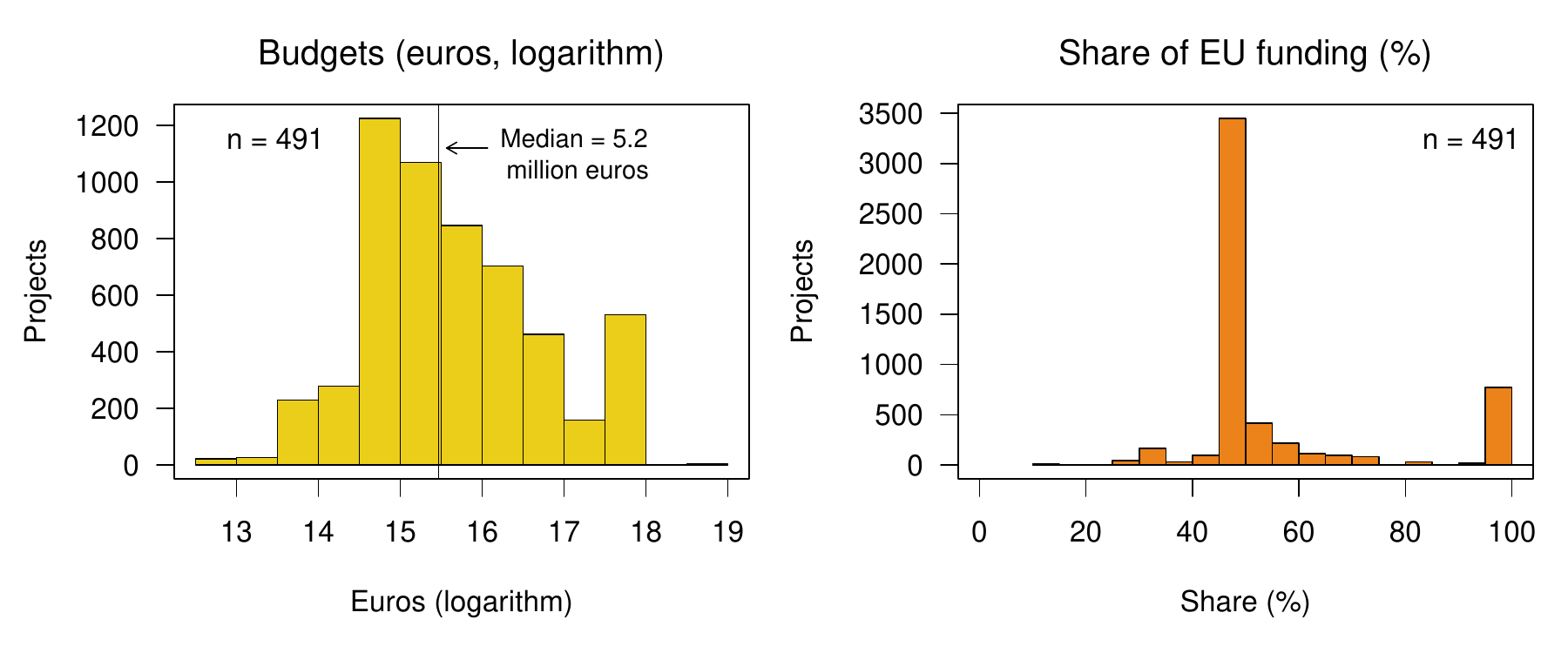}
\caption{Budgets of Projects Funded by the DEP}
\label{fig: budgets}
\end{figure}

The concrete money involved can be elaborated by first taking a brief look at
the actual financial amounts granted through the DEP funding instrument. Thus,
Fig.~\ref{fig: budgets} displays two essential plots about financing. As can be
seen from the left-hand side plot, the median cutoff point about five million
euros separates the DEP-funded projects quite well, although the empirical
distribution does not fully resemble the normal distribution. Regarding the
empirical distribution's tails, there are a few projects that have received only
less than a million euros and a few projects that have received over sixty
million euros. As will be elaborated later on, there is a specific, strategic
reason for this unequal distribution of funds in terms of the smaller and larger
projects funded through the DEP instrument.

Regarding the right-hand side plot, it is evident that a ``fifty-fifty
principle'' has often been quite closely followed in the DEP, meaning that
national consortia have funded a half and the EU the other half. With respect to
the former, it is impossible to say how the national, non-EU funding has been
arranged, but, nevertheless, it is worth remarking a recent criticism that some
projects have failed at attracting private sector investments~\citep{ECA24}. In
any case, the principle is important to emphasize because it indicates a
willingness of the member states to take the DEP's objectives seriously. With
fully funded EU development and innovation projects, there is a risk that funds
granted would go to somewhere else, including even stakeholders and places
associated with corruption. This point notwithstanding, it remains unclear why
some projects have been fully funded by the DEP instrument. For instance, some
national innovation hubs have been entirely funded by the EU, whereas in others
national funding has also been required. The EU's cohesion policy provides a
plausible hypothesis, but further research is required to provide an explanation
beyond~speculation.\footnote{~Regarding cohesion policy, it can be remarked that
among the evaluation criteria for funding applications are geographic balance and
the geographic digital divide in the union; see Article 20(c) in Regulation (EU)
2021/694.}

\begin{figure}[th!b]
\centering
\includegraphics[width=\linewidth, height=12cm]{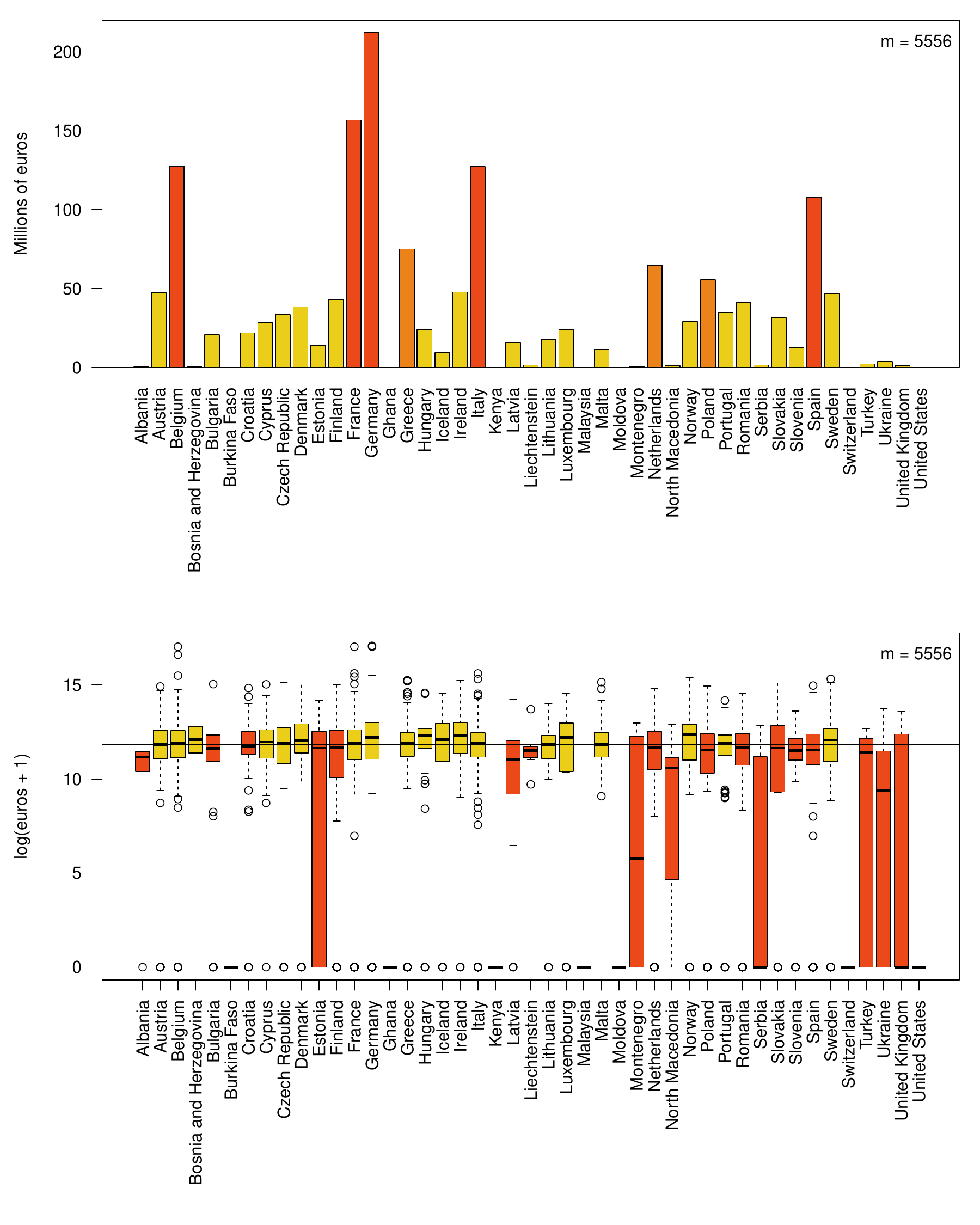}
\caption{EU Funding Granted to Organizations According to Their National Origins}
\label{fig: budgets countries}
\end{figure}

Distribution of financial resources across the member states is always a hot
political topic in the EU. With this truism in mind, Fig.~\ref{fig: budgets
  countries} displays two organization-level plots on the financial amounts
granted by the EU across the national origins of organizations participating in
consortia behind the projects funded through the DEP financial instrument. As
can be seen from the upper plot, the largest member states---Germany, France,
and Italy---have received the most funding in absolute terms. That said, also
Belgium has received considerable financial resources, especially when keeping
the country's size in mind. A~partial explanation relates to the fact that many
important European associations are headquarted in Belgium, among them EIT
Digital that is an important coordination body also in terms of the Horizon
Europe programme.\footnote{~\url{https://www.eitdigital.eu/}} Furthermore,
Herfindahl's classical concentration index---calculated by squaring the funding
shares of each country and summing the result---is as low as
$0.06$.\footnote{~There is a long-standing uncertainty and a controversy
regarding this classical and simple concentration metric. It was supposedly
first published by Orris C. Herfindahl in his unpublished dissertation from
1950~\citep{Hirschman64}.} Such a low value is commonly interpreted to indicate
almost perfect or at least fierce competition between
participants~\citep{Djolov13}. Regardless whether or not competition is a good
term in the present context, an analogous conclusion can be drawn from the lower
plot in Fig.~\ref{fig: budgets countries}. Although there are outliers, the
medians across the member states of the EU are quite close to the overall median
across all countries, as visualized by the horizontal line. Many of the tails
below the overall median, as colored in a darker red, are explained by the fact
that some countries not affiliated with the EU have taken part in consortia but
not received any EU funds. These observations notwithstanding, a point can be
raised that perhaps also the DEP would benefit from a wider political
deliberation. While some of the non-EU countries having received EU funding,
such as Ukraine, may well receive support from the general public, there are
also some a little more controversial cases present. This point serves to remind
that the EU funding granted through the DEP financing instrument is taxpayers'
money.

\subsection*{Consortia}

The previous point is also evident in terms of \text{consortia---or} a lack
thereof. In particular, about ten percent of all projects funded through the DEP
have involved only a single participating organization. While some of these
outliers are understandable and justifiable, such as is the case with the
establishment of national NCCs, other single-organization funding cases are
generally vague. The examples include single cyber security companies, single
individual banks, a single central bank, single individual hospitals, some of
which are private, and generally single private sector
companies.\footnote{~101100648, 101100725, 101101426, 101127710, 101127844,
101128107, 101128109, 101145725, 101145847, 101145867, 101145884, 101145866,
101158662, and 101145877.} It remains generally unclear what were the rationales
for funding such single-organization projects. While they may possess commercial
potential, they seem to generally contradict the overall pan-European focus and
the establishment of innovation hubs, including potential incubators through
which private sector money could have been possibly obtained.\footnote{~Article
20(d) in Regulation (EU) 2021/694 also emphasizes a trans-European dimension as
an evaluation criterion.} Even then, the few cases addressing individual
national banks and individual national hospitals improbably possess a commercial
potential that could be generalized into a product or a service. A~further
related critical point can be raised about some outlying consortia through which
EU funds have been granted for subsidiaries of large multinational companies,
among them Accenture, Cisco, and Microsoft.\footnote{~101158834.}

\begin{figure}[th!b]
\centering
\includegraphics[width=\linewidth, height=4.5cm]{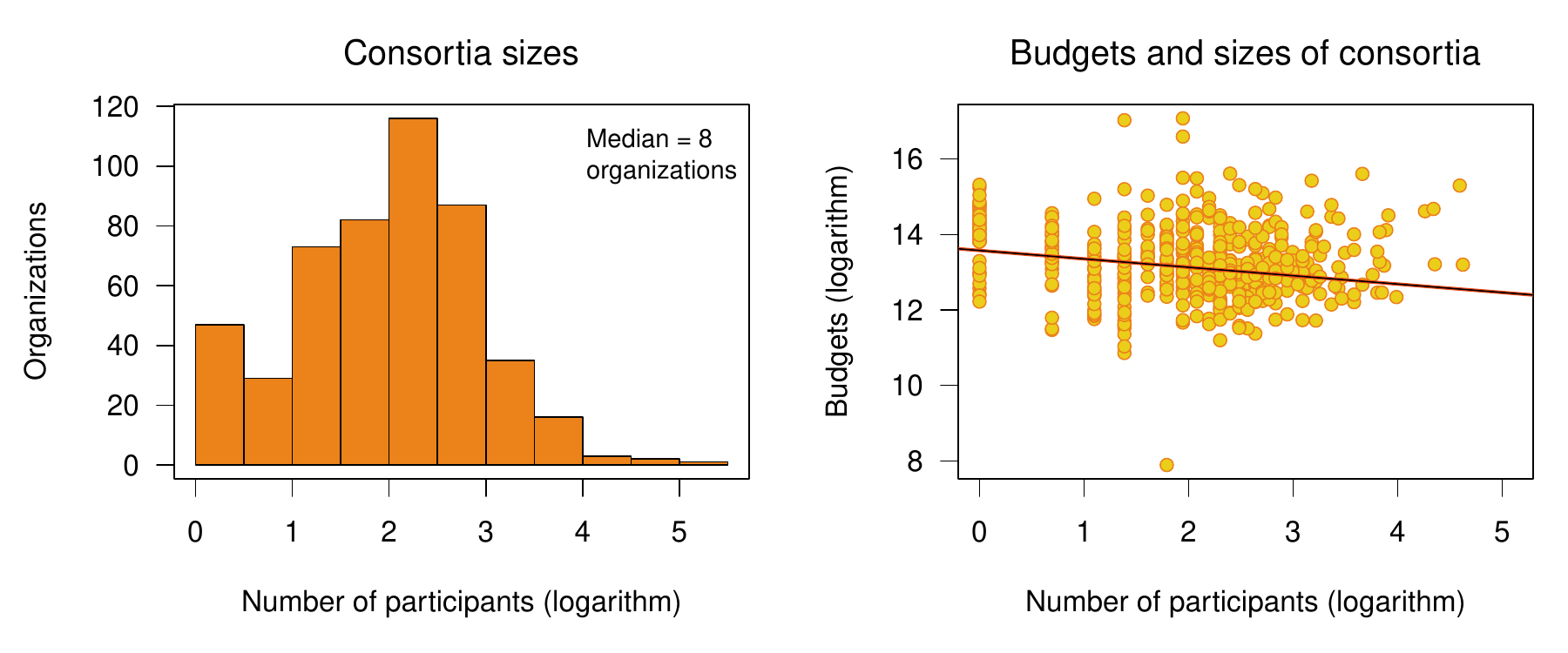}
\caption{Sizes of Consortia}
\label{fig: consortia}
\end{figure}

\begin{figure}[th!b]
\centering
\includegraphics[width=\linewidth, height=10cm]{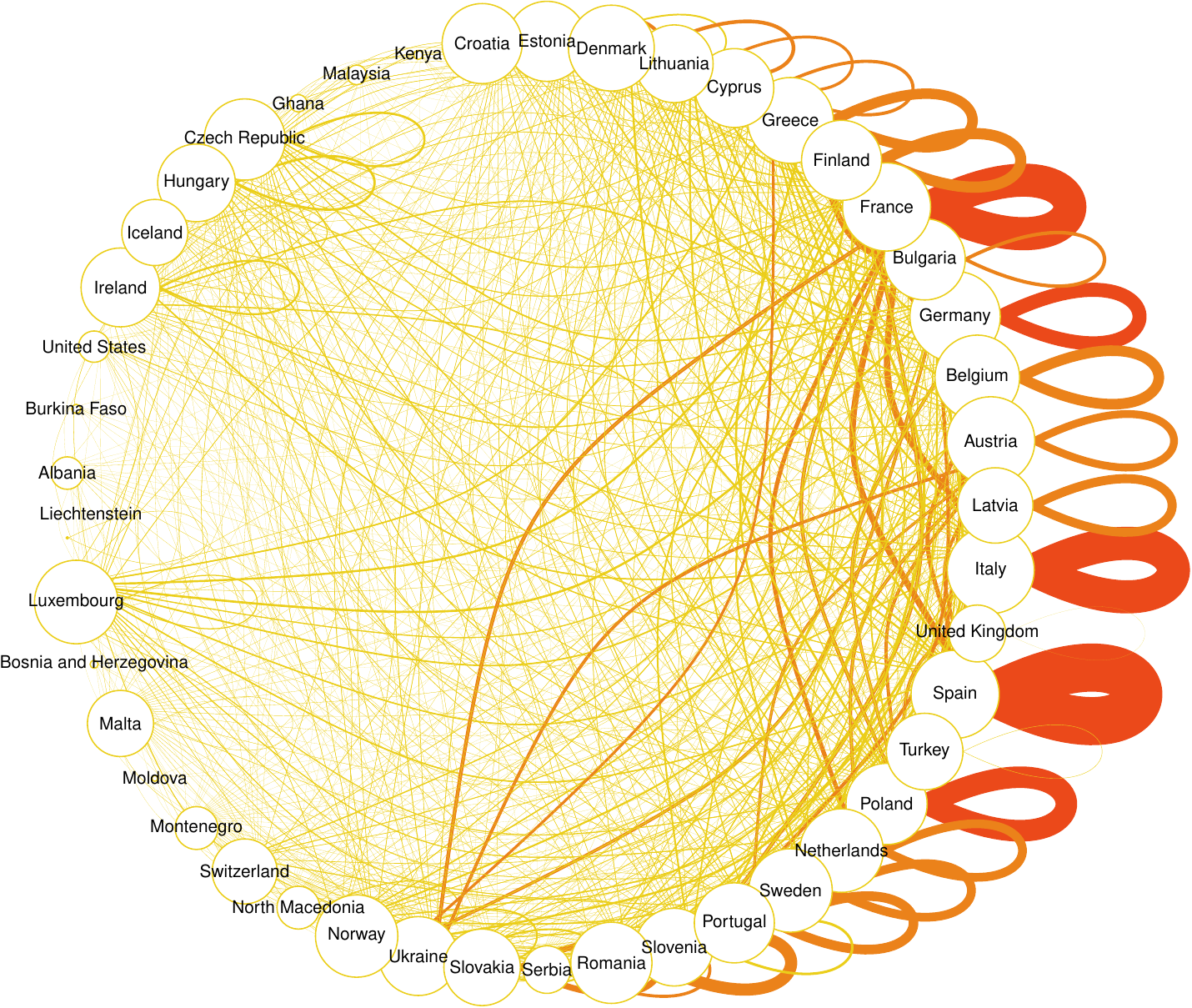}
\caption{Consortia in Terms of National Origins of Participating Organizations}
\label{fig: countries graph}
\end{figure}

Regarding consortia more generally, Fig.~\ref{fig: consortia} displays the sizes
of participating consortia in terms of their organizational memberships. The
left-hand side plot resembles a normal distribution; there are many projects
involving about three to twenty organizations, a few projects involving only one
or two organizations, and then a few particularly large consortia. The
right-hand side plot indicates that there is only a weak correlation between
consortia sizes and their budgets. It is also worth more generally remarking
that bigger is not necessarily always better in the context of consortia;
indeed, one project noted that its ``strong partnership involves a limited
number of actors for better efficiency''.\footnote{~101091448.}

With respect to the particularly large consortia, there are two extreme outliers
involving over a hundred participating organizations. Both are about the EU's
so-called digital wallet, which is essentially about digital identities and
their verification. One is about piloting digital wallet implementations in the
member states.\footnote{~101102655.} The other project is using the already
established ideas and implementations for the so-called eIDAS law, Regulation
(EU) 910/2014, in the education and healthcare sectors; the project's goal is to
have portable digital documents for educational credentials and the European
health insurance card.\footnote{~101102611.} Although not as large in terms of
consortia sizes, it is worth mentioning that there are two additional projects
dealing with digital wallets, including a consortium for building a consortium
around digital wallets.\footnote{~101102744.} The other one is investigating new
payment options through the European digital wallet.\footnote{~101102740.} In
general, these projects are important to emphasize because the digital wallet is
an important part of the EU's grand strategy for digitalization. Although not
yet enacted, there is also a law proposal, identified as COM/2021/281 final, for
the European digital wallet and digital identities in general. According to the proposal's impact assessment, the wallet is expected to have substantial economic benefits.

Furthermore, the geospatial structures of consortia are visualized in
Fig.~\ref{fig: countries graph}. Again, the colors and widths of edges reflect
edge weights; the darker a color and the wider a line, the stronger a
connection. Unlike the previous graph representation in Fig.~\ref{fig: topics
  graph}, this consortia graph includes self-loops from a vertex to itself. Many
such self-loops imply that many countries with a same national origin
participated in a given consortium. The graph visualization is arranged such
that the countries with large amounts of self-loops are on the right-hand
side.

With these clarifying notes about the visualization, it can be concluded that
particularly Spanish, French, Italian, Polish, and, to a lesser extent, German
organizations have collaborated in consortia involving many other Spanish,
French, Italian, Polish, and German organizations. In addition to many results
about the so-called homophily tendency in graphs~\citep[e.g.,][]{Ingram07,
  Wholey93}, a~similar observation has been made previously in the European HPC
domain~\citep{Ruohonen24GJ}. Geographic concentration has been observed to be
present also in the Horizon Europe programme and its
predecessor~\citep{Fernandez19, Kosztyan24, Sekerci23}. Furthermore, geography
has been observed to influence the decisions of European SMEs to adopt new
digital technologies~\citep{Holl24}. Although the present self-loop observation
might relate to the sizes of these countries, a similar result applies also to
many (but not all) smaller member states who too have often collaborated with
their national comrades. However, it is difficult to speculate what might
explain these geospatial results, but they probably have something to do with
the EU itself, its governance and funding instruments, and national interests
and preferences of the member states. Here, it is worth further mentioning that
also \citet{Draghi24} recommended better coordination in the innovation and
technology development domains. Finally, a brief note about clustering. With the
graph's self-loops included, the three clustering algorithms indicate as many as
18, 20, and 34 clusters, meaning that many of the countries with heavy
self-loops cluster into their own groups. When excluding the self-loops, the
amounts are two, four, and nineteen clusters. Only two to four clusters
generally reflect the graph's overall density.

\subsection*{Regression Results}

\begin{figure}[p!]
\centering
\includegraphics[width=\linewidth, height=20cm]{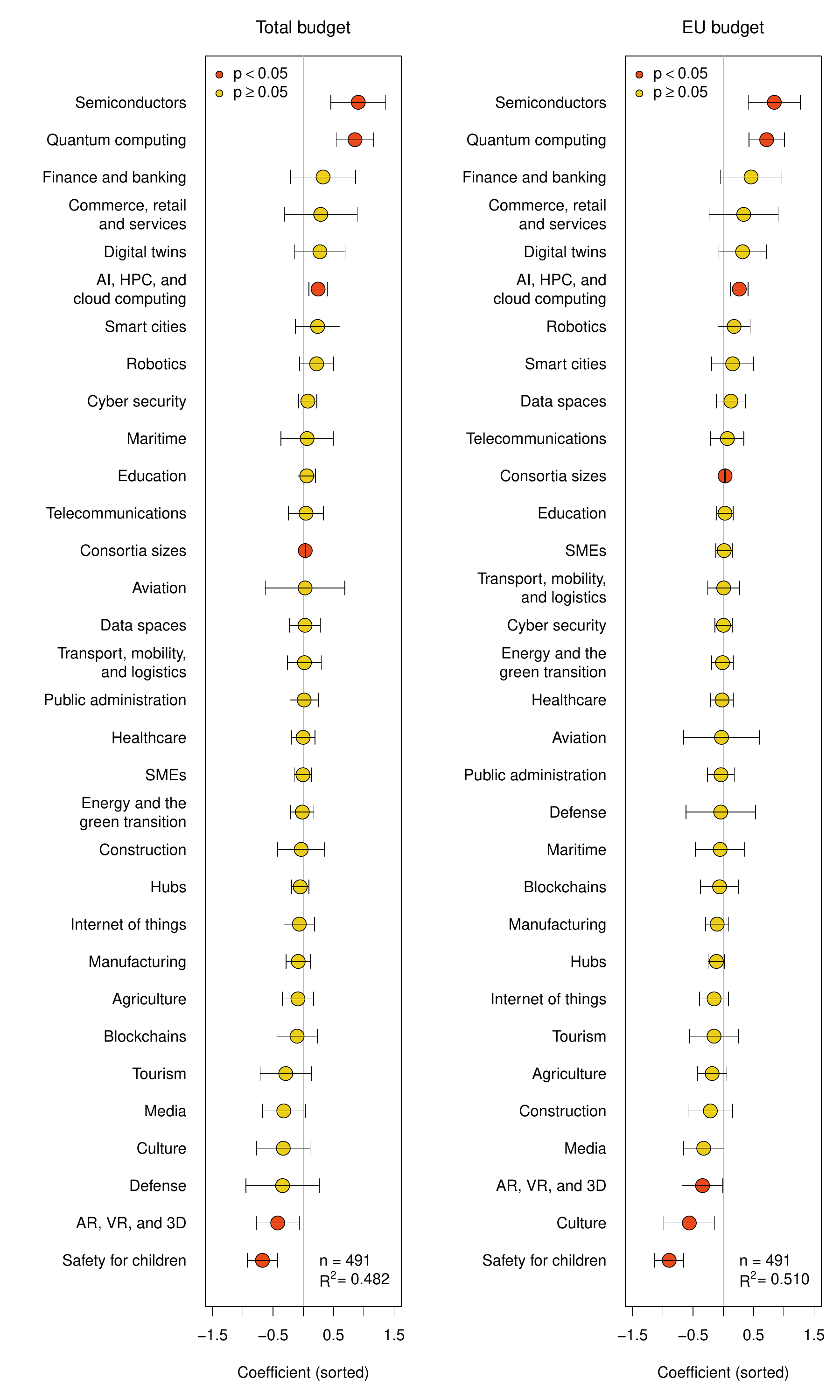}
\caption{Regression Results from the Project-Level Models}
\label{fig: reg project}
\end{figure}

The ordinary least squares regressions serve well to end the empirical
exposition. Thus, the results from the two project-level regressions are shown
in Fig.~\ref{fig: reg project}. To recall: the dependent variables are the
logarithms of the total budgets and EU funds granted, and the independent
variables include the technology domains and economic sectors together with the
consortia sizes. The first observation to draw from the results is that overall
statistical performance is quite good; the coefficients of determination are
close to zero point five, meaning that about a half of the total variances is
explained by the models. The second observation is that the coefficient
estimates are very close to each other between the two models. However, the
third observation is that only a few of the coefficients are statistically
significant at the conventional threshold. This observation can be seen also
from the 95\%-level confidence intervals illustrated with the horizontal
lines. Among the statistically significant coefficients in both models are the
dummy variables for the semiconductor sector and the quantum computing
technology domain. These coefficients are also the largest in terms of
magnitudes. Also the dummy variables for the AI/HPC domain are statistically
significant, although their magnitudes are only small. As already hinted by the
previous Fig.~\ref{fig: countries graph}, the same applies to the consortia
sizes. Among the coefficients with negative signs are the dummy variables for
the AR/VR/3D domain and the culture sector together with the projects addressing
the safety of children online. Finally, it can be noted that basic diagnostics
for both regression models indicate no substantial problems; even the normality
assumption seems relatively unproblematic according to a visual inspection of
quantile-quantile plots of the residuals.

In contrast, the organization-level regression model does not yield good
statistical performance; the coefficient of determination is only $0.12$. None
of the dummy variables for the national origins of participating organizations
stand out in terms of their magnitudes or statistical significance. In general,
this result reflects the lower plot in Fig.~\ref{fig: budgets countries}. It can
be concluded that geography provides only limited explanatory power for
explaining the amounts of EU funding granted.

To return to the project-level results, the semiconductor sector and the quantum
computing domain deserve brief additional comments. In general, it is not
surprising that the coefficients for these two stand out in terms of their
positive signs and magnitudes in Fig.~\ref{fig: reg project}; both are related
to hardware and technology infrastructures, which are typically expensive to
establish and develop. With respect to semiconductors, it is relevant to note
that the DEP was aligned in 2023 with Regulation (EU) 2023/1781, also known as
the Chips Act. The effects of this alignment are visible also in the dataset
observed; there are projects seeking to build alliances in the European
semiconductor sector as well as a few projects designing and developing new
educational programs in higher education.

Then, regarding quantum computing, it can be noted that there are many
DEP-funded national projects piloting and testing the use of new
quantum-specific cryptographic algorithms and related solutions in networking
contexts, including deployments in metropolitan areas. These projects are
further coordinated through the EuroQCI initiative for building quantum
communication infrastructures. It can be also remarked that there are European
        industry associations for quantum computing, such as QuIC who also participates
in the EuroHPC ecosystem.\footnote{~\url{https://www.euroquic.org/}} Also
education is again involved too. For instance, one project envisions fundamental
changes in the coming decade due to quantum computing, seeking to answer to the
changes and the associated challenges by educating a new generation of ``a
quantum-ready workforce''.\footnote{~101084035.} This quantum education project
is related to the earlier remarks about new master-level programs. The
question of how to teach quantum computing is also under active
research~\citep{Haghparast25}. This project and other educational projects
further reiterate the point about the overlapping nature of the technology
domains, economic sectors, and European societies in general.

\section*{Conclusion and Discussion}

The paper examined the Digital Europe Programme for funding technology
development and innovation projects in the European Union. The focus was on the
projects funded through the DEP. Four research questions were postulated and
empirically examined. Brief answers to these provide a way to summarize the
conclusions reached. Thus, the answer to $\textmd{RQ}_1$ is that the DEP's
strategic goals have been met in a sense that projects addressing cyber
security, innovation hubs, SMEs, AI, HPC, and education stand out in terms of
volume. In terms of economic sectors, projects operating in the energy and
healthcare sectors together with the broadly defined manufacturing sector have
frequently been funded when compared to other sectors. Then, the answer to
$\textmd{RQ}_2$ is that on average equal amounts of national and EU funding have
granted. In other words, in most (but not all) cases the EU has funded a half
and national parties the other half. Regarding $\textmd{RQ}_3$, typically
consortia have involved about three to twenty organizations, although there are
outliers involving only a single participating organization on one hand and
particularly large endeavors with more than a hundred organizations on the other
hand. However, the amounts of funding granted do not correlate well the
consortia sizes. In terms of geographic composition, there has been a visible
homophily tendency, meaning that many national organizations have collaborated
particularly intensively with organizations from the same country. With respect
to the last $\textmd{RQ}_4$, information about the technology domains and
economic sectors alone provides enough material for rather good statistical
performance on explaining the amounts of funding granted. As for the individual
domains and sectors, semiconductors and quantum computing have received larger
funding amounts compared to the rest.

The answer to $\textmd{RQ}_1$ is enough to conclude that the DEP's explicitly
defined objectives, as specified in Article~3 of Regulation (EU) 2021/694, have
more or less been met during the early roll-out of the funding instrument. In
other words, the DEP is on the right track with respect to this part of its
\textit{ex~ante} objectives. However, the early evaluation check conducted
revealed also some unintended consequences, such as the occasional, outlying
funding granted for multinational companies. When going through the regulation's
recitals, some omissions and uncertainties seem to be also present. Among other
things, recital~10 emphasizes that data protection, digital rights, and
fundamental rights should be guaranteed. Although many of the projects,
including those dealing with AI in particular, have emphasized compliance and
ethics, it is impossible to evaluate how well these non-technical requirements
have been taken into account in practice. Then, recital 54 notes that open
source solutions should be encouraged. Although some of the projects may have
opened their sources or plan to do so, none of the project descriptions
explicitly mention open source, whether software or something else. This point
supports recent arguments that funding of open source software has often but not
entirely been overlooked by political
decision-makers~\citep{Ruohonen24JSS}. Finally, recital 24 remarks that the DEP
and its objectives should be evaluated also against those specified the InvestEU
Programme established via Regulation (EU) 2021/523. The DEP seems to largely
satisfy the additional objectives, including the promotion of research,
innovation, and digitalization, improving social resilience (through education),
and helping SMEs. However, there are also other objectives, such as the
promotion of territorial cohesion and improvement of capital markets, that are
more difficult to evaluate and assess. These provide also a good motivation for
pinpointing further research topics.


What about the theoretical motivation of differentiation? An answer to this
question is more difficult to give. On one hand, the conclusion reached for
$\textmd{RQ}_1$ could be used to support and justify reasoning with the
differentiation concept. That is, the DEP's strategic framing and its selective
prioritization approach indicate a degree of differentiation. While all are
important, culture and media do not weigh as much as quantum computing and
semiconductors according to the DEP's decision-makers, and so forth and so
on. On the other hand, it could be also argued that despite the instrument's
strategic goals and associated differentiation, the DEP is still largely about
``all things digital''. To this end, it might be contemplated whether a sharper
focus and further differentiation would be beneficial. A similar point applies
to the hubs established, \text{many---but} again not \text{all---of} which seem
to lack specialization and a strategic differentiation from the hundreds of
other hubs in Europe. A comparable point can be raised also in terms of economic
sectors. Here, the NIS2 directive is a good example because mild criticism has
been levied that ``all things critical'', in turn, have been considered in the
directive~\citep[cf.][]{Ruohonen24ISJGP}. This point has manifested itself also
through the DEP because there are many projects addressing cyber security in
some particular sectors. Because everything is increasingly digital and many
things are considered as critical, a logical conclusion would be that more and
more economic sectors would justifiably need public funding through the DEP or
other financing instruments in the future. These arguments and contemplations
motivate five points about further relevant research topics.

First, the note about early perspectives in the introduction requires a
revisit. That is to say, further evaluation research is needed once the DEP has
ran for a sufficient amount of time. It may be that also the fund's strategic
focus will slightly change in the future. The innovation hubs would be a good
example; it may be likely that not many new hubs will be established in the
future.

Second, even with the ECCC and NCCs, which are backed by law, a critical
argument has been raised that too little guidance has been provided on the
practical details for building European cyber security
communities~\citep{Ferreira20}. A~similar point applies to the innovation hubs
established through the DEP. In particular, there is no EU-level coordination
body for the hubs, and in many cases it seems that even national coordination
and oversight are loose. That said, flexibility and freedom may also be seen in
a positive light; after all, without them, innovation is hindered---if not
doomed altogether. In addition, regarding the EU-level, a critical argument has
been raised that the ECCC is an example of increasing supranationalism and its
close cousin, federalism~\citep{Suciu22}. Thus: given the political turn toward
conservatism and nationalism in many member states, there are also political
risks involved in increasing coordination and synchronization through EU-level
administration. A further related point is that Regulation (EU) 2021/694 may
have locked too strictly the DEP's objectives. According to previous
evaluations, synergies between funding instruments have sometimes occurred when a
scope has been defined broadly, as has been the case with many societal grand
challenges, such as the green transition~\citep{Doussineau21}. The point applies
also to the hubs. Although it is impossible to deduce what the hubs are
currently doing in practice, the associated projects for their establishment
have often justified their funding applications by meticulously listing what the
law desires. This point is also a limitation for the present work. In other
words, it may be that cyber security, hubs, SMEs, and AI/HPC in particular are
biased upwards because they are all something that the funding body wants to
hear. To address this potential limitation, further evaluation work is required
about actual results of the projects and their correspondence with the
funding~applications.

Third, a good related question for further research would be whether the DEP has
contributed to the heterogeneity of the European innovation landscape, as has
sometimes been argued to be a risk for the Horizon Europe
programme~\citep{Veugelers15}. In other words, merely making funding decisions
at the EU-level do not necessarily translate into coherence at the development
and implementation level, whether in terms of technological harmony or
geographic cohesion. This assertion entails further research on the governance
and coordination aspects behind the DEP and the EU's other related funding
instruments. The topic is also related to the hubs established throughout
Europe. As has been recently argued~\citep{Colovic25}, the activities performed
by them and their performance at facilitating innovation are not
well-understood. To this end, it would be interesting to know whether a similar
homophily tendency is present among the hubs and their stakeholders. As it
stands, it seems that cross-border collaboration is limited between the hubs,
but it may well be that homophily tendencies are present also at the regional
level. If so, a subsequent question would be whether or not such tendencies are
a good thing or a bad thing in terms of technology development and
innovation. With this point, the Rubicon is also crossed in a sense that
cohesion policy is still involved even with differentiation at the EU's
supranational level. Furthermore, as has been pointed out with respect to the
defense and security domains~\citep{Rieker24}, merely providing funding for the
establishment of hubs and other collaborative endeavors does not guarantee the
livelihood and sustainability of the establishments initially funded. With this
point in mind, further research is required also on the potential funding
required to keep the hubs operational in their day-to-day activities. Arguably,
it would seem to be a bad strategic choice to fund something that might not
survive without further public funding.

Fourth, more contributions are needed to deduce about whether the DEP has
managed to attract also some private sector investments.\footnote{~Integration
and strengthening of the EU's fragmented capital markets are also specified as an
objective in Article~3(e) of Regulation (EU) 2021/523.} The topic is important
because a lack of large-scale venture capital has often been seen as a problem
for the European innovation system in general~\citep{Fernandez19}. The
coordination between public sector funding organizations and private venture
capitalists has also been seen as a problem in a sense that the former have
often merely thrown money at the latter with a hope of ``breeding
unicorns''~\citep{Alami24}. In the present context the topic could be initially
probed by examining the structure of national funding granted for consortia
taking part in the DEP. If also the national funding mostly comes from national
budgets of the member states, it would seem that money is merely transferred
from a public purse to another public purse via the fifty-fifty percent
principle discussed earlier.

\enlargethispage{2cm} 

Fifth and last, further work is required to investigate the alleged synergy
benefits between the DEP and the EU's other funding instruments, including the
Horizon Europe programme in particular. Regulation (EU) 2021/694 enumerates such
synergies in its Annex III, but it seems that only a few projects have justified
their applications with these. That said, there are outliers; there is a
DEP-funded project seeking to develop further the ideas enacted in two Horizon
Europe projects.\footnote{~101101322.} It would be worthwhile to study whether
something could be learned from such outliers. The topic is not easy because it
involves the always difficult demarcation between research, development,
innovation, and science and technology in general. Nevertheless, it might be
interesting and even relevant to sketch new blue sky ideas. One such idea might
relate to joint, possibly simultaneous applications to the Horizon Europe
programme and the DEP. Having some ideas, even if preliminary, about the science
involved might benefit the latter, while the former might benefit from knowing
about practical implementations and adaptations, again even in terms of sketches
and prototypes.

\bibliographystyle{apalike}

\begin{thebibliography}{}

\bibitem[Alami, 2024]{Alami24}
Alami, I. (2024).
\newblock {S}tate {P}roperty, {V}enture {C}apital and the {U}rbanisation of
  {S}tate {C}apitalism.
\newblock {\em Dialogues in Human Geography}, (Published online in May):1--5.

\bibitem[Bachtr\"ogler-Unger and Doussineau, 2021]{Doussineau21}
Bachtr\"ogler-Unger, J. and Doussineau, M. (2021).
\newblock {E}xploring {S}ynergies {B}etween {EU} {C}ohesion {P}olicy and
  {H}orizon 2020 {F}unding across {E}uropean {R}egions : {A}n {A}nalysis of
  {R}egional {F}unding {C}oncentration on {K}ey {E}nabling {T}echnologies and
  {S}ocietal {G}rand {C}hallenges.
\newblock {E}uropean {C}ommission, {JRC} Technical Reports, the Joint Research
  Centre (JRC), Luxembourg, Publications Office of the European Union,
  available online in January 2025:
  \url{https://publications.jrc.ec.europa.eu/repository/handle/JRC123485}.

\bibitem[Brodny and Tutak, 2022]{Brodny22}
Brodny, J. and Tutak, M. (2022).
\newblock {A}nalyzing the {L}evel of {D}igitalization {A}mong the {E}nterprises
  of the {E}uropean {U}nion {M}ember {S}tates and {T}heir {I}mpact on
  {E}conomic {G}rowth.
\newblock {\em Journal of Open Innovation: Technology, Market, and Complexity},
  8:1--29.

\bibitem[Colovic et~al., 2025]{Colovic25}
Colovic, A., Caloffi, A., Rossi, F., and Russo, M. (2025).
\newblock {I}nstitutionalising the {D}igital {T}ransition: {T}he {R}ole of
  {D}igital {I}nnovation {I}ntermediaries.
\newblock {\em Research Policy}, 54:105146.

\bibitem[Cs\'ardi et~al., 2024]{igraph}
Cs\'ardi, G., Nepusz, T., Traag, V., Horv\'at, S., Zanini, F., Noom, D.,
  M\"uller, K., Salmon, M., and Antonov, M. (2024).
\newblock igraph: {N}etwork {A}nalysis and {V}isualization.
\newblock {R} package version 2.1.2, available online in December 2024:
  \url{https://cran.r-project.org/web/packages/igraph/index.html}.

\bibitem[Dewi and Nugrahani, 2024]{Dewi24}
Dewi, A.~S. and Nugrahani, H. S.~D. (2024).
\newblock {S}trengthening {EU} {C}yber {R}esilience: {A} {C}ritical {A}nalysis
  of the {C}yber {S}olidarity {A}ct's {L}egislative {F}ramework.
\newblock {\em Islamic World and Politics}, 8(2):182--195.

\bibitem[Djolov, 2013]{Djolov13}
Djolov, G. (2013).
\newblock {T}he {H}erfindahl-{H}irschman {I}ndex as a {D}ecision {G}uide to
  {B}usiness {C}oncentration: {A} {S}tatistical {E}xploration.
\newblock {\em Journal of Economic and Social Measurement}, 38:201--227.

\bibitem[Draghi, 2024]{Draghi24}
Draghi, M. (2024).
\newblock {T}he {F}uture of {E}uropean {C}ompetitiveness.
\newblock {P}art~{A}: {A} {C}ompetitiveness {S}trategy for {E}urope, available
  online in January 2025:
  \url{https://commission.europa.eu/document/download/97e481fd-2dc3-412d-be4c-f152a8232961_en}.

\bibitem[{ECA}, 2024]{ECA24}
{ECA} (2024).
\newblock {A}nnual {R}eport on {EU} {J}oint {U}ndertakings for the {F}inancial
  {Y}ear 2023.
\newblock {E}uropean {C}ourt of {A}uditors {(ECA)}, Luxembourg, Publications
  Office of the European Union, available online in December 2024:
  \url{https://www.eca.europa.eu/ECAPublications/SAR-JUS-2023/SAR-JUS-2023_EN.pdf}.

\bibitem[{European Commission}, 2021]{EC21}
{European Commission} (2021).
\newblock {ANNEX} to the {C}ommission {I}mplementing {D}ecision on the
  {F}inancing of the {D}igital {E}urope {P}rogramme and {A}doption of the
  {M}ultiannual {W}ork {P}rogramme -- {E}uropean {D}igital {I}nnovation {H}ubs
  for 2021 -- 2023.
\newblock {C(2021)} 7911 final, available online in January 2024:
  \url{https://ec.europa.eu/newsroom/dae/redirection/document/80907}.

\bibitem[{European Commission}, 2022a]{EC22a}
{European Commission} (2022a).
\newblock {C}ommission {S}taff {W}orking {D}ocument on {C}ommon {E}uropean
  {D}ata {S}paces.
\newblock SWD(2022) 45 final, available online in December 2024:
  \url{https://ec.europa.eu/newsroom/dae/redirection/document/83562}.

\bibitem[{European Commission}, 2022b]{EC22b}
{European Commission} (2022b).
\newblock {E}uropean {E}nterprise {S}urvey on the {U}se of {T}echnologies
  {B}ased on {A}rtificial {I}ntelligence.
\newblock Luxembourg, Publications Office of the European Union, available
  online in January 2025:
  \url{https://op.europa.eu/en/publication-detail/-/publication/f089bbae-f0b0-11ea-991b-01aa75ed71a1/language-en}.

\bibitem[{European Commission}, 2024a]{EC24a}
{European Commission} (2024a).
\newblock {EU} {F}unded {P}rojects: {D}igital {E}urope {P}rogramme {(DEP)}.
\newblock {EU} {F}unding \& {T}enders {P}ortal, available online in December
  2024:
  \url{https://ec.europa.eu/info/funding-tenders/opportunities/portal/screen/opportunities/projects-results?isExactMatch=true&frameworkProgramme=43152860}.

\bibitem[{European Commission}, 2024b]{EC24b}
{European Commission} (2024b).
\newblock {P}roposal for a {J}oint {E}mployment {R}eport from the {C}ommission
  and the {C}ouncil.
\newblock {COM}(2024) 701 final. Available online in December 2024:
  \url{https://commission.europa.eu/document/download/23d334e4-f4b8-480c-bdcf-111a40c0c4f0_en?filename=COM_2024_701_1_EN.pdf}.

\bibitem[Fern\'{a}ndez et~al., 2019]{Fernandez19}
Fern\'{a}ndez, S.~G., Kubus, R., and {P\'{e}rez-I\~{n}igo}, J.~M. (2019).
\newblock {I}nnovation {E}cosystems in the {EU}: {P}olicy {E}volution and
  {H}orizon {E}urope {P}roposal {C}ase {S}tudy (the {A}ctors' {P}erspective).
\newblock {\em Sustainability}, 11:1--25.

\bibitem[Ferreira et~al., 2022]{Ferreira20}
Ferreira, A., {von Wintzingerode}, C., M\"ullmann, D., Benzekri, A., Cros,
  P.-H., D\"ohmann, I.~S., and Prochilo, E. (2022).
\newblock {G}overnance {F}oundations for the {E}uropean {C}ybersecurity
  {C}ommunity.
\newblock In {\em Proceedings of the First International Workshop on Digital
  Sovereignty in Cyber Security: New Challenges in Future Vision
  (CyberSec4Europe 2022)}, pages 132--148, Venice. Springer.

\bibitem[Flohr, 2008]{Flohr08}
Flohr, T. (2008).
\newblock {A}ssessing the {Q}uality of {Q}uality {G}ate {R}eference
  {P}rocesses.
\newblock In {\em Proceedings of the Third IFIP TC2 Central and East-European
  Conference on Software Engineering Techniques (CEE-SET 2008)}, pages
  207--217, Brno. Springer.

\bibitem[Haghparast et~al., 2025]{Haghparast25}
Haghparast, M., Moguel, E., Garcia-Alonso, J., Mikkonen, T., and Murillo, J.~M.
  (2025).
\newblock {I}nnovative {A}pproaches to {T}eaching {Q}uantum {C}omputer
  {P}rogramming and {Q}uantum {S}oftware {E}ngineering.
\newblock {A}rchived manuscript, available online in January 2025:
  \url{https://arxiv.org/abs/2501.01446}.

\bibitem[Hirschman, 1964]{Hirschman64}
Hirschman, A.~O. (1964).
\newblock {T}he {P}aternity of an {I}ndex.
\newblock {\em The American Economic Review}, 54(5):761--762.

\bibitem[Holl and Rama, 2024]{Holl24}
Holl, A. and Rama, R. (2024).
\newblock {S}patial {P}atterns and {D}rivers of {SME} {D}igitalisation.
\newblock {\em Journal of the Knowledge Economy}, 15:5625--5649.

\bibitem[Ingram and Morris, 2007]{Ingram07}
Ingram, P. and Morris, M.~W. (2007).
\newblock {D}o {P}eople {M}ix at {M}ixers? {S}tructure, {H}omophily, and the
  ``{L}ife of the {P}arty''.
\newblock {\em Administrative Science Quarterly}, 52:558--585.

\bibitem[Koszty\'an et~al., 2024]{Kosztyan24}
Koszty\'an, Z.~T., Kir\'ly, F., Katona, A.~I., Csizmadia, T., and
  Feh\'erv\"olgyi, B. (2024).
\newblock {A}nalysis and {P}rediction of the {H}orizon 2020 {R}\&{D}\&{I}
  {C}ollaboration {N}etwork.
\newblock {\em Expert Systems with Applications}, 255(Part B):124417.

\bibitem[Krogstie, 2024]{Krogstie24}
Krogstie, J. (2024).
\newblock {A}nvendelse av kunstig intelligens (ki) i {N}orge i norsk offentlig
  sektor 2024.
\newblock {A}rchived manuscript, available online in January 2025:
  \url{https://arxiv.org/abs/2412.19273}.

\bibitem[Letta, 2024]{Letta24}
Letta, E. (2024).
\newblock {M}uch {M}ore {T}han a {M}arket.
\newblock {A} {R}eport {P}repared for the {E}uropean {C}ommission, available
  onine in January 2025:
  \url{https://www.consilium.europa.eu/media/ny3j24sm/much-more-than-a-market-report-by-enrico-letta.pdf}.

\bibitem[Lindgren et~al., 2020]{Lindgren20}
Lindgren, B.-M., Lundman, B., and Graneheim, U.~H. (2020).
\newblock {A}bstraction and {I}nterpretation {D}uring the {Q}ualitative
  {C}ontent {A}nalysis {P}rocess.
\newblock {\em International Journal of Nursing Studies}, 108:103632.

\bibitem[Molica, 2025]{Molica25}
Molica, F. (2025).
\newblock {R}eassessing {C}ohesion {P}olicy {T}hrough the {L}ens of the {N}ew
  {EU} {I}ndustrial {P}olicy.
\newblock {\em Journal of Common Market Studies}, 63(1):302--319.

\bibitem[Newman, 2006]{Newman06}
Newman, M. E.~J. (2006).
\newblock {F}inding {C}ommunity {S}tructure {U}sing the {E}igenvectors of
  {M}atrices.
\newblock {\em Physical Review E}, 74:036104.

\bibitem[Newman and Girvan, 2004]{NewmanGirvan04}
Newman, M. E.~J. and Girvan, M. (2004).
\newblock {F}inding and {E}valuating {C}ommunity {S}tructure in {N}etworks.
\newblock {\em Physical Review E}, 69:026113.

\bibitem[Rieker and Giske, 2024]{Rieker24}
Rieker, P. and Giske, M. T.~E. (2024).
\newblock {\em {E}uropean {A}ctorness in a {S}hifting {G}eopolitical {O}rder:
  {E}uropean {S}trategic Autonomy {T}hrough {D}ifferentiated {I}ntegration}.
\newblock Palgrave Macmillan, Cham.

\bibitem[Rosvall and Bergstrom, 2007]{Rosvall07}
Rosvall, M. and Bergstrom, C.~T. (2007).
\newblock {M}aps of {R}andom {W}alks on {C}omplex {N}etworks {R}eveal
  {C}ommunity {S}tructure.
\newblock {\em PNAS}, 105(4):1118--1123.

\bibitem[Ruohonen, 2024a]{Ruohonen24ISJGP}
Ruohonen, J. (2024a).
\newblock {A} {S}ystematic {L}iterature {R}eview on the {NIS2} {D}irective.
\newblock {A}rchived manuscript, available online:
  \url{https://arxiv.org/abs/2412.08084}.

\bibitem[Ruohonen, 2024b]{Ruohonen24GJ}
Ruohonen, J. (2024b).
\newblock {G}eospatial {I}nsights on the {E}uro{HPC} {S}upercomputing
  {E}cosystem.
\newblock {A}rchived manuscript, available online:
  \url{https://doi.org/10.31219/osf.io/z94f2}.

\bibitem[Ruohonen, 2024c]{Ruohonen24I3E}
Ruohonen, J. (2024c).
\newblock {T}he {I}ncoherency {R}isk in the {EU}'s {N}ew {C}yber {S}ecurity
  {P}olicies.
\newblock In {\em Proceedings of the 23rd IFIP Conference on e-Business,
  e-Services, and e-Society (I3E 2024)}, pages 284--295, Heerlen. Springer.

\bibitem[Ruohonen et~al., 2024]{Ruohonen24JSS}
Ruohonen, J., Choudhary, G., and Alami, A. (2024).
\newblock {A}n {O}verview of {C}yber {S}ecurity {F}unding for {O}pen {S}ource
  {S}oftware.
\newblock {A}rchived manuscript, available online:
  \url{https://arxiv.org/abs/2412.05887}.

\bibitem[Schoenefeld and Jordan, 2017]{Schoenefeld17}
Schoenefeld, J. and Jordan, A. (2017).
\newblock {G}overning {P}olicy {E}valuation? {T}owards a {N}ew {T}ypology.
\newblock {\em Evaluation}, 23(3):274--293.

\bibitem[Sekerci and Alp, 2023]{Sekerci23}
Sekerci, D. and Alp, S. (2023).
\newblock {I}nvestigation of {E}uropean {U}nion {H}orizon 2020 {I}nformation
  and {C}ommunication {T}echnology {P}rojects with the {S}ocial {N}etwork
  {A}nalysis {M}ethod.
\newblock {\em Engineering, Technology \& Applied Science Research},
  13(4):11182--11190.

\bibitem[Serini, 2024]{Serini24}
Serini, F. (2024).
\newblock {C}ollective {C}yber {S}ituational {A}wareness in {EU}. {A}
  {P}olitical {P}roject of {D}ifficult {L}egal {R}ealisation?
\newblock {\em Computer Law \& Security Review}, 55:106055.

\bibitem[Siddi et~al., 2022]{Siddi22}
Siddi, M., Karjalainen, T., and Jokela, J. (2022).
\newblock {D}ifferentiated {C}ooperation in the {EU}'s {F}oreign and {S}ecurity
  {P}olicy: {E}ffectiveness, {A}ccountability, {L}egitimacy.
\newblock {\em The International Spectator}, 57(1):107--123.

\bibitem[Smismans, 2015]{Smismans15}
Smismans, S. (2015).
\newblock {P}olicy {E}valuation in the {EU}: {T}he {C}hallenges of {L}inking
  {E}x {A}nte and {E}x {P}ost {A}ppraisal.
\newblock {\em European Journal of Risk Regulation}, 6(1):6--26.

\bibitem[Suciu and Cirjan, 2022]{Suciu22}
Suciu, S. and Cirjan, A.-L. (2022).
\newblock {T}he {E}uropean {C}ybersecurity {C}ompetence {C}entre -- {O}ne
  {M}ore {S}tep {T}owards {S}upranationalism.
\newblock {\em Perspective Politice}, 15(1--2):57--73.

\bibitem[Teixeira and {Tavares-Lehmann}, 2022]{Teixeira22}
Teixeira, J.~E. and {Tavares-Lehmann}, A. T. C.~P. (2022).
\newblock {I}ndustry 4.0 in the {E}uropean {U}nion: {P}olicies and {N}ational
  {S}trategies.
\newblock {\em Technological Forecasting \& Social Change}, 180:121664.

\bibitem[Vaismoradi and Snelgrove, 2019]{Vaismoradi19}
Vaismoradi, M. and Snelgrove, S. (2019).
\newblock {T}heme in {Q}ualitative {C}ontent {A}nalysis and {T}hematic
  {A}nalysis.
\newblock {\em FQS}, 20(3):1--14.

\bibitem[Veugelers and Cincera, 2015]{Veugelers15}
Veugelers, R. and Cincera, M. (2015).
\newblock {H}ow to {T}urn on the {I}nnovation {G}rowth {M}achine in {E}urope.
\newblock {\em Intereconomics}, 50(1):4--30.

\bibitem[{von der Leyen}, 2024]{vonderLeyen24}
{von der Leyen}, U. (2024).
\newblock {S}tatement at the {E}uropean {P}arliament {P}lenary by {P}resident
  {U}rsula von der {L}eyen, {C}andidate for a {S}econd {M}andate 2024--2029.
\newblock {E}uropean {C}ommission, available online in January 2025:
  \url{https://ec.europa.eu/commission/presscorner/detail/en/STATEMENT_24_3871}.

\bibitem[Weck, 2024]{CESinfo24}
Weck, T. (2024).
\newblock {EU} {C}ompetitiveness at a {C}rossroads: {W}hy the {D}raghi {R}eport
  {F}alls {S}hort, and the {EU} {T}reaties {O}ffer a {S}olution.
\newblock {\em EconPol Forum}, 25(6):26--29.
\newblock {A}vailable online in January 2025:
  \url{https://www.cesifo.org/DocDL/econpol-forum-2024-6-weck-eu-regulation.pdf}.

\bibitem[Wholey and Huonker, 1993]{Wholey93}
Wholey, D.~R. and Huonker, J.~W. (1993).
\newblock {E}ffects of {G}eneralism and {N}iche {O}verlap on {N}etwork
  {L}inkages {A}mong {Y}outh {S}ervice {A}gencies.
\newblock {\em Academy of Management Journal}, 36(2):349--371.

\end{thebibliography}

\end{document}